\documentstyle[preprint,prb,eqsecnum,aps]{revtex}
\input{epsf}
\input epsf.sty

\begin{document}
\draft
\title{
An accurate effective action for `baby' to `adult' skyrmions
}
\author{Kyungsun Moon and Kieran Mullen}
\address{
Department of Physics and Astronomy, University of Oklahoma, Norman,
OK~~73019\\
}
\date{\today}
\maketitle

{\tightenlines
\begin{abstract}
Starting with a Chern-Simons theory,
we derive an effective action for 
interacting quantum Hall skyrmions 
that takes into account both large-distance physics and  
short-distance details as well.
We numerically calculate the classical static skyrmion profile from  
this action and find excellent agreement 
with other,  microscopic calculations over 
a wide  range of skyrmion sizes including the experimentally 
relevant one.   This implies that the essential physics of this regime
might be captured by
a continuum classical model rather than resorting to more microscopic approaches.
We also show that the skyrmion energy 
closely follows the formula suggested earlier by Sondhi {\em et al.}
for a broad parameter range of interest as well.
\end{abstract}
}

\pacs{PACS numbers: 74.60Ec, 74.75.+t, 75.10.-b}

\narrowtext

\section{Introduction} 
Quantum Hall 
systems\cite{GirvinMacdBook} near filling factor $\nu=1/(2n+1)$ 
should manifest a 
charged topologically stable object called a 
skyrmion.\cite{LeeandKane,Sondhi1,MacdFertig,Kmoon,Belavin,WuandJain,Mrasolt,Rajaraman,Xie}
Various experiments
have shown  strong evidence for the existence of this exotic object in two 
dimensional electron systems.
\cite{Barrett,Eisenstein,Aifer} In particular the ground state 
of the quantum Hall system near  filling factor  $\nu=1$ is believed 
to be a manybody state of weakly interacting skyrmions.
\cite{MacdFertig,SkyrmCry} 

Just as the thermodynamics of certain superconductors
can be well-described by their vortex degrees of freedom  alone,  
skyrmions, the defects of the incompressible quantum Hall liquid,
may well describe the essential physics of the  quantum Hall system near
$\nu=1$.
The thermodynamics of   vortex systems can be derived from the
phenomenological Ginzburg-Landau (GL) description of superconductivity.
In traditional BCS-theory the spin-singlet Cooper-pairing mechanism  
removes the electronic spin degrees of freedom from the 
dynamics. Hence vortices 
have no  spin structure,\cite{comment2} and are entirely
described by their integer charge.
In contrast, the quantum Hall skyrmion is a topological spin-texture made 
of an intricate pattern of electron-spin orientations.  It possesses internal
degrees of freedom associated with this texture that allow for 
novel dynamics.\cite{SkyrmCry}  

In analogy to GL theory in the superconductivity, 
the Chern-Simons theory of quantum Hall systems was derived by
Zhang {\em et al.}\cite{Zhang} and subsequently extended 
by Lee and Kane\cite{LeeandKane} 
to describe the possible spin-unpolarized quantum Hall liquid.
In superconductivity,
it has been very useful to obtain a phenomenological
theory of vortices based on GL theory.\cite{jose} 
A similar theory for skyrmions should be invaluable for understanding
transport, phase transitions, and spatial ordering.\cite{SkyrmCry}

In this paper, 
starting from the Chern-Simons theory\cite{LeeandKane,Zhang}  
we derive an action for the many skyrmion system
which takes into account both large-distance physics\cite{comment1} 
and relatively short-distance details as well.
Using this action we perfrom numerical studies of 
the classical skyrmion solution. 
Recently Abolfath {\em et al.}\cite{Abolfath} 
pointed out 
for small skyrmions, {\em  i.e.} those typical in  GaAs 
samples,\cite{Barrett,MacdFertig,Fertig} 
the minimal effective field theory does not give a good quantitative agreement
with the Hartree-Fock  or exact-diagonalization study. 
We find that the various properties of a static skyrmion solution
obtained from the above action exhibit an excellent agreement with the   
microscopic study for `baby' skyrmions (textures containing 
from two to about $10$ flipped spins) and larger.  
This implies that current experiments can be
well modelled by a continuum theory, obviating the need for
the more involved microscopic approaches.

In section II, we briefly summarize the Chern-Simons description
of quantum Hall effect and then derive an action for many skyrmion
system.   In section III, the classical skyrmion profile 
is numerically calculated. The various properties of a static skyrmion 
solution are compared with the microscopic study.
We show that the skyrmion energy follows the formula 
suggested earlier by Sondhi {\em et al.} very accurately
for a broad parameter range.
In section IV, we conclude with a summary.

\section{Chern-Simons theory}

We briefly summarize the Chern-Simons description of quantum Hall effect
which was subsequently extended by Lee and Kane 
in order to incorporate a possible spin-unpolarized quantum Hall liquid.
\cite{Zhang,LeeandKane}
In the bosonic Chern-Simons theory\cite{Zhang}, an electron
is viewed as a composite object of a boson and a flux tube carrying
an odd-multiple  of flux quanta $\phi_0=h/e$ 
attached via a Chern-Simons term, which correctly ensures fermionic 
statistics for the electron.
We begin with the effective Chern-Simons Lagrangian introduced by 
Lee and Kane\cite{LeeandKane} 
\begin{eqnarray}
{\cal L}[\Psi_\sigma,a_\mu]&=& {\bar \Psi}_\sigma (\partial_0 - i a_0) 
\Psi_\sigma
+ {1\over 2m} \left| ({\frac {\nabla}{i}}-{\bf a}-{\bf A}_{\rm ex}) 
\Psi_\sigma \right|^2 \nonumber\\
&+& {1\over 2}\int d{\bf r}^\prime 
\left({\bar \Psi}_\sigma({\bf r})
 \Psi_\sigma({\bf r})
-{\bar \rho}\right)\,\, V(|{\bf r}-{\bf r}^\prime|) 
\,\,\left({\bar \Psi}_\sigma({\bf r}^\prime)
\Psi_\sigma({\bf r}^\prime)
-{\bar \rho}\right) + {\cal L}_{\rm cs} 
\end {eqnarray}
where $\Psi_\sigma$ represents a bosonic field with spin 
$\sigma=\pm {1\over 2}$, $m$ the effective mass of boson, ${\bf A}_{\rm ex}$  
a vector potential for the external magnetic field, ${\bar \rho}$
a mean density of boson, and $V(r)=e^2/\epsilon r$   
Coulomb interaction between electrons. 
We will use the convention of units where $\hbar=c=e=1$.
The Chern-Simons term ${\cal L}_{\rm cs}$ can be written as
\begin{equation}
{\cal L}_{\rm cs} = \frac {i}{4\pi\alpha}\epsilon_{\rm \mu\nu\lambda}
a_\mu\partial_\nu a_\lambda 
\end {equation}
where $a_\mu$ is a statistical gauge field with $\mu=0,1,2$, 
($1$ and $2$ are spatial
indices, $0$ is time)  and we impose 
a Coulomb gauge $\nabla\cdot{\bf a}=0$. 
The variable $\alpha$ is taken to be an odd integer, $\alpha=2n+1$, 
in order to describe a fermionic system.  

In order to separate the amplitude and the spin degrees of freedom from
the bosonic field $\Psi_\sigma$, we introduce the   
$CP^1$-field $z_\sigma$, which satisfies $|z_\sigma|=1$ and
is related to $\Psi_\sigma$ by
\begin{equation}
\Psi_\sigma=\sqrt{J_0} \,\, z_\sigma(x)  
\end{equation}
where $J_0$ represents the boson density.  
The Lagrangian in terms of the fields $J_0$ and $z_\sigma$ is given by
\begin{eqnarray}
{\cal L}[z_\sigma,a_\mu] &=& {1\over 2}\int d{\bf r}^\prime 
\,\, \left(J_0({\bf r})-{\bar \rho}\right)
\,\, V(|{\bf r}-{\bf r}^\prime|) \,\, \left(J_0({\bf r}^\prime)-{\bar \rho}\right)
+i J_0 \,\, ({\bar z}_\sigma \frac {\partial_0}{i} z_\sigma -a_0 ) \nonumber\\  
&+&\frac {\kappa}{2} \left|{\bar z}_\sigma \frac {\nabla}{i} z_\sigma 
-{\bf a}-{\bf A}_{\rm ex}\right|^2
+ {\cal L}_{\rm NL\sigma} + {\cal L}_{\rm cs} 
\end{eqnarray}
where $\kappa={\bar \rho}/m$ and the gradient term in $J_0({\bf r})$ 
is neglected.\cite{Mstone} 
The local spin orientation ${\bf m}$ is related to 
the field $z_\sigma$ via ${\bf m}={\bar z_\alpha}
{\vec{\sigma}}_{\alpha\beta} z_\beta$, so that the static non-linear sigma model 
(${\rm NL}\sigma$) term can be written by
\begin{equation}
{\cal L}_{{\rm NL}\sigma}={\kappa\over 8} (\nabla {\bf m})^2.
\end{equation}
In order to decouple the quartic term, one introduces a Hubbard-Stratonovich
field which represents a bosonic current ${\bf J}(x)$. After integrating out
the statistical gauge field $a_\mu$, one obtains the following Lagrangian:
\cite{Mstone}
\begin{eqnarray}
{\cal L}[z_\sigma,A_\mu] &=& {1\over 2}
\int d{\bf r}^\prime \left(J_0({\bf r})-{\bar \rho}\right) 
\,\, V(|{\bf r}-{\bf r}^\prime|) \,\, \left(J_0({\bf r}^\prime)-{\bar \rho}\right)
+{1\over 2\kappa} |{\bf J}|^2 + {\cal L}_{\rm NL\sigma}  \nonumber\\
&+& i (A_\mu + A^{(0)}_\mu)\,\,  J^s_\mu - {i\over 2} \alpha A_\mu
\,\, (J_\mu-J^{(0)}_\mu) + {i\over 2\pi} A_0\,\,  (B-2\pi\alpha{\bar \rho})
\label{Stone}
\end{eqnarray}
where $J_\mu=J^{(0)}_\mu + {1\over 2\pi} \epsilon_{\rm \mu\nu\lambda}
\partial_\nu A_\lambda $ and $ J^{(0)}_\mu= ({\bar \rho},0,0)=
{1\over 2\pi} \epsilon_{\rm \mu\nu\lambda}\partial_\nu A^{(0)}_\lambda$.
Here we used the fact that the bosonic current and density satisfies
a continuity equation $\partial_\mu J_\mu=0$. 
The skyrmion {\em three}-vector $J^s_\mu$ can be written in terms of ${\bf m}$:
\begin{equation}
J^s_{\mu} ={1\over 8\pi\alpha}
\epsilon_{\mu\nu\lambda} (\partial_\nu {\bf m}\times
\partial_\lambda {\bf m})\cdot {\bf m} 
\end{equation} 
where we explicitly put the factor of $1/\alpha$ in the definition
of $J^s_{\mu}$.  The $J^s_0(x)$ is the topological charge density of the
spin texture, which is proportional to the electronic charge density.

Zhang {\em et al.} have shown that when external magnetic field
is tuned so that the number of flux quanta
is commensurate with the mean boson density, 
($B=2\pi\alpha{\bar \rho}$, so that, $\nu=1/(2n+1)$), the bosons condense
and form a superfluid\cite{Zhang}. The bosonic superfluidity in the
Chern-Simons theory implies that a quantum Hall effect occurs in the 
corresponding two-dimensional electron system. 
The ground state has been shown to be 
a fully spin-polarized quantum ferromagnet.   
By noticing that the dynamics of quantum Hall system with a spin-polarized 
ground state will follow that of a quantum ferromagnet and  that the
skyrmion is a charged object of the system\cite{LeeandKane}, 
Sondhi {\em et al.} proposed a phenomenological action,  
which is valid for the long-wavelength and small-frequency limit.

However we can explicitly integrate out the bosonic field 
in Eq.(\ref{Stone}) 
and derive an action for skyrmions which takes fully into account
the short- and long-distance physics.   A similar exercise is
standard in the GL theory for vortices.\cite{jose,stern}
We proceed by integrating out the bosonic field $A_\mu$ with a Coulomb gauge 
condition,
$\nabla\cdot {\bf A}=0$.
In order to impose the Coulomb gauge, we introduce an auxilliary
field $\lambda(x)$ and introduce an additional term 
$i\lambda(x) \nabla\cdot{\bf A}$ into Eq.(\ref{Stone}).
Since the action is still quadratic in the field $A_\mu$ and 
diagonal in the frequency-momentum space, one can exactly
integrate out the bosonic field $A_\mu$.
The integration of the ${\bf k}=0$ mode leads to the following
relations between the skyrmion density $n_s$ and the external 
magnetic field $B$, {\em i.e.} $n_s=(2\pi\alpha{\bar \rho}-B)/(2\pi)$.

After lengthy but straightforward calculations, we finally obtain
the following action 
\begin{eqnarray}
{\cal S}_E[{\bf m}]&=&{1\over 2}\sum_{{\bf k},\omega} \frac 
{\kappa \alpha^2 V(k)}{P(k,\omega)} \,\,
|J_0^s|^2 + {1\over 2}\sum_{{\bf k},\omega}
\frac {\alpha^2}{P(k,\omega)} |{\bf J}^s|^2 
+ \int d\tau \int d{\bf r} \,\, {\cal L}_{{\rm NL}\sigma} + 
\int d\tau \int d{\bf r} \,\, {\cal L}_z\nonumber\\ 
&+& i\alpha\sum_{{\bf k},\omega} {\bf A^{(0)}}(-k)\cdot {\bf J}^s (k)  
-\sum_{{\bf k},\omega} \frac {2\pi\kappa\alpha^3}
{P(k,\omega)k^2} \,\, J_0^s(-k){\hat z}\cdot 
{\bf k}\times {\bf J}^s(k) \nonumber\\
&+& \sum_{{\bf k},\omega} \frac {Q(\omega)\alpha^2} 
{P^2 (k,\omega)+ Q^2(\omega)}\left\{ \frac {{\bf k}\cdot {\bf J}^s(-k) 
\,\, \left({\bf k}\times {\bf J}^s(k)\right)\cdot {\hat z}}{k^2} - {1\over 2}
{\bf J}^s(-k)\times {\bf J}^s(k)\cdot {\hat z}\right\}
\label{grand}
\end{eqnarray}
where $k$ stands for $(\omega,{\bf k})$, $P(k,\omega)=\kappa\alpha^2 
+ V(k) k^2/(4\pi^2) + 
\omega^2/(4\pi^2\kappa)$, and  
$Q(\omega)=\alpha\omega/(2\pi)$ and the Zeeman term ${\cal L}_z$ is given by 
\begin{equation}
{\cal L}_z = \frac {t}{2\pi\ell^2} \,\,\left(1-m_z({\bf r})\right)
\end{equation}
where the magnetic length $\ell=(\hbar c /|e|B)^{1/2}$ and 
$t=(1/2) g\mu_B B$.
In the above action, the first term represents a charge-density 
interaction between skyrmions including the self-energy contribution.
The function $V(k)$ is the Fourier transform of the Coulomb interaction;
due to the additional term $V(k) k^2/(4\pi^2)\propto k$ in the denominator,
the interaction is modified by the short-range fluctuations of the gauge 
field  
from a Coulombic one $\sim 1/r$ to $\ln(1/r)$ 
at short-distances.
The next term represents the kinetic energy for skyrmion.
In the limit of long-wavelengths and small-frequencies, we obtain 
the following action:
\begin{eqnarray}
{\cal S}_E [{\bf m}]&=& {1\over 2}\sum_{{\bf k},\omega} V(k) \,\, 
|J_0^s|^2 + \sum_{{\bf k},\omega}
{1\over 2\kappa}  \,\,|{\bf J}^s|^2 + \int d\tau \int 
d{\bf r}  \,\,{\cal L}_{{\rm NL}\sigma} 
+ \int d\tau \int d{\bf r}  \,\,{\cal L}_z\nonumber\\
&+&i\alpha\sum_{{\bf k},\omega} {\bf A^{(0)}}(-k)\cdot {\bf J}^s (k) 
-\sum_{{\bf k},\omega} \frac {2\pi\alpha}
{k^2} \,\, J_0^s(-k){\hat z}\cdot
{\bf k}\times {\bf J}^s(k) .
\label{Lagra}
\end{eqnarray}
The term $i\alpha{\bf A^{(0)}}({\bf r}_i)\cdot {\bf J}^s_i$
indicates that the skyrmion views the original boson as 
a magnetic flux tube\cite{Mstone}.
The last term in Eq.(\ref{Lagra}) contains the 
exchange-statistics of skyrmion. 
It can be re-written into the more suggestive form 
$i{\bf A}_{sk}\cdot {\bf J}^s$ where 
\begin{equation}
{\bf \nabla}\times {\bf A}_{sk}=2\pi\alpha J^s_0 .
\end{equation}
In order to see the exchange-statistics of a skyrmion, 
suppose that all the other
skyrmions are at rest while one moves around the static 
skyrmion configuration.
When a skyrmion traverses around a closed loop, this term generates
 a phase proportional to the number of skyrmions  enclosed in the loop:
\begin{equation}
\int d{\bf r} \int_0^T dt \,\, {\bf A}_{sk}\cdot {\bf J}^s=2\pi\alpha q_{\rm sk}
\int_{S} d{\bf r} \,\, J^s_0
\end{equation}
where $S$ stands for the 
space enclosed by the closed skyrmion loop.
Using the fact that the skyrmion charge $q_{\rm sk}$ is equal to
$e/\alpha$, one can show that skyrmion picks up a phase 
$(2\pi/\alpha) N_{\rm enc}$,  
where $N_{\rm enc}$ is the number of skyrmions enclosed.
Now consider a process which exchanges two skyrmions:
in the rest frame of one of the skyrmions, the exchange corresponds 
to the other skyrmion moving about the first in a half circle; 
and hence it picks up a phase $\pi/\alpha$.
Since $\alpha$ is $(2n+1)$, 
the statistical phase of a skyrmion is $\pi/(2n+1)$.
For $n=0$, skyrmion is a fermion, while for $n\ne 0$, 
it's an anyon\cite{Mstone,Kun}.

\section{Numerical solution of classical skyrmion}

We calculate the classical skyrmion solution with varying Zeeman energy
and make a comparison to the microscopic result obtained by Hartree-Fock
 and exact diagonalization studies.
We begin with the static energy functional $E[{\bf m}]$ derived from
Eq.(\ref{grand}) 
\begin{equation}
E[{\bf m}]={1\over 2}\sum_{{\bf k}} \frac
{\kappa \alpha^2 V(k)}{\kappa \alpha^2+V(k) k^2/(4\pi^2)}
|J_0^s|^2 + {\kappa\over 8}\int d{\bf r} (\nabla {\bf m})^2  
+ \frac {t}{2\pi\ell^2} \int d{\bf r} (1-m_z({\bf r})) .
\end{equation}
Note that the charge-density interaction changes from 
a Coulombic one $\sim 1/r$ to $\ln(1/r)$ at short-distances.
Since the skyrmion size is determined by balancing the Zeeman energy
and the Coulomb interaction, the size of skyrmion will be reduced
from the estimates of the minimal field theory\cite{Abolfath}.

By using the fact that the skyrmion solution is azimuthally-symmetric,
we choose the form 
${\bf m}=\left(\sin\theta(r)\cos\phi,\sin\theta(r)\sin\phi,\cos\theta(r)\right)$.
In order to solve for the classical skyrmion profile, we extremize the 
energy functional $E[{\bf m}]$ with respect to $\theta(r)$ and 
obtain the following equation\cite{Abolfath} 
\begin{equation}
{1\over r}{\partial\over \partial r}\left(r{\partial\theta\over \partial r}
\right)-\frac {\sin 2\theta}{2r^2}-\sqrt{32\over \pi}{\tilde g}
\sin \theta  
+{1\over \pi}
\sqrt{{2\over\pi}}\frac {\sin\theta}{r}f(r) = 0
\label{DifE}
\end{equation} 
where ${\tilde g}=2t/(e^2/\epsilon\ell)$ and $f(r)$ is given by  
\begin{equation}
f(r)=\int dr^\prime \left({d\over dr} U(r,r^\prime)\right) \sin\theta(r^\prime)
{d\theta\over dr^\prime} 
\end{equation}
where the azimuthally-averaged interaction potential,  $U(r,r^\prime)$, is
given by $U(r,r^\prime)=4\int_0^{\pi/2}d\phi \,\, 
{\tilde V}(\sqrt{(r-r^\prime)^2+4rr^\prime
\sin^2\phi})$ with ${\tilde V}(r)=\int_0^\infty dq \,\, J_0(qr)/(1+bq)$,  
$b=2\sqrt{2/\pi}$, and $J_0$ the Bessel function of the zeroth order.  
If instead we set $b=0$, we recover the
 case of a pure Coulomb interaction, and  then Eq.(\ref{DifE}) 
agrees with the one obtained by 
Abolfath {\em et al.}\cite{Abolfath}.
We have used the fact that $\kappa$ is equal 
to $4\rho_s$ where  
$\rho_s=e^2/(16\sqrt{2\pi}\epsilon \ell)$ as shown by 
Sondhi {\em et al.}\cite{Sondhi1}  
and by Moon {\em et al.}\cite{Kmoon}.
We impose the boundary 
conditions: $\theta(r=0)=\pi$ and $\theta(r\rightarrow \infty)=0$. 

A brief explanation of how the non-local term $f(r)$ is handled is
 appropriate.
In order to calculate the function $f(r)$, we first need to obtain the
explicit form of a modified skyrmion interaction ${\tilde V}(r)$ following
the integration
over momenta $q$ by standard numerical integration methods.
The interaction potential  ${\tilde V}(r)$ varies as $1/r$ for $r\gg b$ and
$(1/b)\,\ln (1+b/r)$ for $r\ll b$.  By virtue of this asymptotic
behaviour, we accurately approximate ${\tilde V}(r)$ by  
$(1/b)\ln (1+b/r) + 1/(a_0+a_1 r + a_2 r^2 + a_3 r^3 )$, where
the best fit parameters are $a_0=13.222, a_1=6.158, a_2=1.223$, and  
$a_3=0.0004$.
After differentiating  ${\tilde V}(\sqrt{(r-r^\prime)^2+4rr^\prime
\sin^2\phi})$ with respect to $r$,
the function $R(r,r^\prime)\equiv dU(r,r^\prime)/dr$
can be written as follows
\begin{equation}
R(r,r^\prime)=4 \int_0^{\pi/2} d\phi \frac {r-r^\prime + 2 r^\prime
\sin^2\phi}{(r-r^\prime)^2 + 4 r r^\prime \sin^2\phi} H(x)
\end{equation}
where $x\equiv\sqrt{(r-r^\prime)^2 + 4 r r^\prime \sin^2\phi}$ 
and $H(x)\equiv x(d{\tilde V}/dx)$ satifies $\lim_{x\to 0} H(x)=-1/b$.
We can now perform the integration over $\phi$.
By noticing that for $r\cong r^\prime$, then $R(r,r^\prime)$ 
as a function of $r'$ 
asymptotically approaches  a step-function $-(\pi/4r)
{\rm sgn}(r^\prime-r)$, we se that $R(r,r^\prime)$ 
can be decomposed into a regular and a discontinuous part:
\begin{equation}
R(r,r^\prime)={\pi\over 2r} \Theta(r-r^\prime) +
R_{\rm reg} (r,r^\prime)\end{equation}
where $\Theta(x)$ is the Heaviside step function 
and the $R_{\rm reg}$ is a smooth, continuous function.
Then $f(r)$ can be decomposed as well yielding
\begin{equation}
f(r)={\pi\over 2r} ( 1 + \cos\theta(r)) + \int_0^\infty dr^\prime R_{\rm reg}
(r,r^\prime) {d\over dr} \cos\theta(r^\prime).
\end{equation}
The integration in the second term can be {\em accurately} done
by  standard numerical methods.

In order to solve the differential equation numerically, we discretized
the equation by $N_0$ segments with uniform
spacing $\Delta r$ in units of the magnetic length $\ell$.
First, the differential equation is solved for a typical experimental value of  
${\tilde g}=0.015$, 
which gives a suitable set of parameters for the number of discretization 
$N_0$ and the spacing $\Delta r$.
The set of parameters are chosen to be $N_0=500$ and 
$\Delta r=0.1$;  the results are not sensitive to  
reasonable choice of $N_0$ and $\Delta r$. 
For other values of ${\tilde g}$, we  re-scale $\Delta r$ by the  
length scale $\xi\propto 1/\sqrt{{\tilde g}}$
set by the Zeeman energy, which controls the size of skyrmion 
: $\Delta r=0.1\sqrt{0.015/{\tilde g}}$.
The differential equation is solved using an iterative method on the finite
number of grid $N_0$. 

We first calculate the number of spin-flips as a function of ${\tilde g}$
which is defined as follows\cite{Abolfath}
\begin{equation}
K={1\over 4\pi\ell^2}\int d{\bf r} (1-\cos\theta(r))-{1\over 2}.
\end{equation}
In Fig.(\ref{SpinFlip}), the solid curve is our field-theoretical result 
for $K\ge 2$, 
which shows a very good agreement with the Hartre-Fock calculation by Fertig 
{\em et al.}\cite{Fertig}. 
Since the Hartree-Fock  and exact diagonalization method 
can not easily calculate the single skyrmion energy due to the long-range 
nature of Coulomb interaction, the energy cost for creating a charge-neutral 
object is calculated by Abolfath {\em et al.}\cite{Abolfath}.
However it has been pointed out by Abolfath {\em et al.} that 
the energy difference $\Delta(K)$ between the skyrmion 
with $K$ and $K+1$ flipped spins can be still obtained.
$\Delta(K)$ corresponds to $2t$ at which an energy level crossing
occurs between a skyrmion with $K$ spin-flips and $K+1$ 
spin-flips.\cite{Abolfath}
Since $K$ is not quantized in the continuum field theory, 
$\Delta(K)$ roughly corresponds to the value of $2t$ 
at the field where the number of flipped spins equals $K+1/2$.
In Fig.(\ref{DeltaK}), $\Delta(K)$ is plotted with respect to $K$.
The solid curves are obtained from our field-theoretical calculation.
Our field-theoretical calculation gives good overall agreement with the
microscopic calculations ranging from  
`baby' skyrmions ($2<K<10$) to full-fledged skyrmions.  
For  `infant' skyrmions, {\em i.e.} $K\le 2$, the quantum fluctuations 
about the classical skyrmion solution will become important\cite{Abolfath}.
Since the number of spin flips in typical experiments with 
GaAs samples\cite{Barrett,MacdFertig,Fertig} is $K\sim 3$, 
our field-theoretical calculation  
produces a reasonable result over the whole parameter range of interest.
For relatively large (`adult') skyrmions, Sondhi {\em et al.} 
have suggested the following 
asymptotic formula for the skyrmion energy\cite{Sondhi1,Sondhi2}
: $E[{\tilde g}]/(4\pi\rho_s) = 1 + A/(4\pi)[{\tilde g} 
\ln({\tilde g})]^{1/3}$ with $A=(3\pi^2/4)(72/\pi)^{1/6}\cong 24.9$.
One can see that in the absence of Zeeman energy, 
{\em i.e.} ${\tilde g}\rightarrow 0$,
the skyrmion energy correctly approaches to the ${\rm NL}\sigma$ model result.
The above formula is {\em not} completely on rigorous footing yet 
as commented by Sondhi {\em et al.} in their recent publication\cite{Sondhi2}.
In Fig.(\ref{Energy}), the skyrmion energy is plotted with respect to $[{\tilde g}
\ln({\tilde g})]^{1/3}$, which shows a nice linear behaviour 
for small ${\tilde g}$.
We estimate the line slope by a linear fit and $A$ is obtained to be   
$24.9 \pm 0.1$. 
Hence we have numerically confirmed that the skyrmion energy follows the
formula quite accurately over a very broad parameter range. For example, 
for typical experimental value of ${\tilde g}=0.015$ where $K\sim 3.7$, 
the energy obtained from the above formula is within $0.2 \%$ of our
numerical estimates.
Since the Hartree-Fock calculation is numerically limited to rather small
skyrmion sizes,\cite{Abolfath} we believe that our field theoretical
calculation incorporating short-distance physics will be a valuable
tool to get a quantitative information over a wide
range of skyrmion sizes.

\section{summary} 
We have obtained an effective Skyrmion action which incorporates short-distance
physics as well as large-distance physics based on the Chern-Simons theory. 
We have numerically calculated the classical skyrmion profile and shown that 
for $K\ge 2$, our field theoretical results exhibit an 
excellent agreement with the microscopic study.
We have also demonstrated that the skyrmion energy very closely follows 
the formula suggested earlier by Sondhi {\em et al.} over 
a broad parameter range of interest. 
We believe that our field theoretical
calculation incorporating short-distance physics will be a valuable
tool to get a quantitative information for skyrmion over a wide parameter  
range of skyrmion sizes with reasonable numerical effort.

\section{acknowledgements}
It is our great pleasure to acknowledge useful conversations with  
S.M. Girvin, C.L. Kane, D.H. Lee, S.L. Sondhi, and M. Stone.
We want to give special thanks to H. Fertig and A.H. MacDonald 
for allowing us to use their data in our paper.
The work was supported by NSF DMR-9502555.  K.~Mullen is partially
supported under an Oklahoma EPSCoR grant via LEPM from
the National Science Foundation.

\begin{figure}
\caption{The number of spin flip as a function of ${\tilde g}$:
The solid curve corresponds to our field theoretical result
and the open circles the Hartree-Fock calculation 
by Fertig {\em et al.}[20].
}
\label{SpinFlip}
\end{figure}

\begin{figure}
\caption{$\Delta(K)$ as a function of the number of spin-flip $K$.
The solid curve is calculated from the field theory.
The filled circle represents the Hartre-Fock data and the open circles
the exact diagonalization result 
by Abolfath {\em et al.}[19].
}
\label{DeltaK}
\end{figure}

\begin{figure}
\caption{The skyrmion energy as a function of $[{\tilde g}
\ln({\tilde g})]^{1/3}$.
The line is a linear fit to our numerical data for small ${\tilde g}$.
}
\label{Energy}
\end{figure}

\end{document}